\documentclass[epj]{svjour}
\usepackage{graphics}
\usepackage{amsmath}
\usepackage{graphicx}

\begin{document}
\title{Configurational statistics of densely and fully packed loops in the negative-weight percolation model}
\author{O. Melchert \and A. K. Hartmann}
\institute{
Institut f\"ur Physik, Universit\"at Oldenburg, 
Carl-von-Ossietzky Strasse, 26111 Oldenburg, Germany} 
\date{Received: date / Revised version: date}
%
\abstract{
By means of numerical simulations we investigate the configurational
properties of  densely and fully packed configurations of loops in the
negative-weight percolation (NWP) model.  In the presented study we
consider $2d$ square, $2d$ honeycomb, $3d$ simple cubic
and $4d$ hypercubic lattice
graphs, where edge weights are drawn from a Gaussian distribution.
For a given realization of the disorder we then compute a
configuration of loops, such  that the configurational energy, given
by the sum of all individual loop weights,  is minimized. For
this purpose, we employ a mapping of the NWP model to
the ``minimum-weight perfect matching problem'' that can be solved 
exactly by
using sophisticated polynomial-time
matching algorithms.  We characterize the loops
via observables similar to those used in percolation  studies and
perform finite-size scaling analyses, up to side length $L\!=\!256$ 
in $2d$, $L\!=\!48$ in $3d$ and $L\!=\!20$ in $4d$ 
(for which we study only some observables), in order to estimate
geometric exponents that characterize the configurations of densely
and fully packed loops. One major result is that the loops
behave like uncorrelated random walks from dimension $d=3$ on, in
contrast to the previously studied behavior at the percolation threshold, where
random-walk behavior is obtained for $d\ge 6$.
} 
\authorrunning{O. Melchert \and A. K. Hartmann}
\titlerunning{Configurational statistics of DPLs and FPLs in the NWP model}
\maketitle
%
\section{Introduction \label{sect:introduction}}

The statistical properties of lattice-path models on graphs,
equipped with quenched disorder, have experienced much attention during the 
last decades.
They have proven to be useful in order to describe line-like quantities as, e.g.,
linear polymers in disordered media \cite{kardar1987,derrida1990,grassberger1993,buldyrev2006}, 
vortices in  high $T_c$ superconductivity \cite{pfeiffer2002,pfeiffer2003},
cosmic strings in the early universe \cite{vachaspati1984,scherrer1986,hindmarsch1995},
and domain-wall excitations in disordered media such as $2d$ spin glasses \cite{cieplak1994,melchert2007} and 
the $2d$ solid-on-solid model \cite{schwarz2009}. 
The precise computation of these paths can often be formulated in terms
of a combinatorial optimization problem and hence might allow for the
application of exact optimization algorithms developed in computer science \cite{schwartz1998,rieger2003,SG2dReview2007}.
So as to analyze the statistical properties of these lattice path models, 
geometric observables and scaling concepts similar to those used in 
percolation theory \cite{stauffer1979,stauffer1994} or other ``string''-bearing
models \cite{allega1990,austin1994} are often applicable. 

The topic of the presented article is the \emph{negative-weight percolation} 
(NWP)  problem \cite{melchert2008,apolo2009,melchert2010a}, 
wherein one considers a regular 
lattice graph with periodic boundary conditions
(BCs) and where adjacent sites are joined by undirected edges. Weights are
assigned to the edges, representing quenched random variables drawn from 
a distribution that allows for edge weights of either sign. 
For a given realization of the disorder, one then computes a 
configuration of loops, i.e.\ closed paths on the lattice graph, 
such that the total sum of the weights assigned to the edges
that build up the loops attains an 
\emph{exact} minimum.
Note that the application  of exact algorithms in contrast
to standard sampling approaches like Monte Carlo simulations
avoids, e.g., equilibration problems. Also, since the algorithms run
in polynomial time, large instances can be solved.
As an additional optimization constraint we impose the condition that 
the loops are not allowed to intersect. 
Consequently, since a loop does neither intersect with itself nor 
with other loops in its neighborhood, it exhibits an ``excluded volume''
quite similar to usual self avoiding walks (SAWs) \cite{stauffer1994}.
The problem of finding these loops can be cast into a minimum-weight
path (MWP) problem, outlined below in sect.\ \ref{sect:model} in more detail.

In previous studies \cite{melchert2008,apolo2009,melchert2010a}, the
details of the weight distribution were further controlled by a tunable disorder 
parameter. As a pivotal observation it was found that, as a function of the disorder parameter,  
the NWP model features a disorder driven, geometric phase transition.
This transition leads from a phase characterized by only ``small'' loops to a phase that also
features ``large'' loops that span the entire lattice along at least one direction.
Regarding these two phases and in the limit of large system sizes, there is a particular 
value of the disorder parameter at which percolating (i.e.\ system spanning) loops appear 
for the first time \cite{melchert2008}.     
Previously, we have investigated the NWP phenomenon for $2d$ lattice graphs \cite{melchert2008}
using finite-size scaling (FSS) analyses, where we characterized the underlying transition by 
means of a whole set of critical exponents. 
Considering different disorder distributions and lattice geometries, the exponents were found 
to be universal in $2d$ and clearly distinct from those describing other percolation phenomena.
In a subsequent study we investigated the effect of dilution on the critical 
properties of the $2d$ NWP phenomenon \cite{apolo2009}. Therefore we performed
FSS analyses to probe critical points along the critical line in the disorder-dilution plane 
that separates domains that allow/disallow system spanning loops. 
One conclusion of that study was that bond dilution changes the universality class 
of the NWP problem. Further we found that, for bond-diluted lattices prepared
at the percolation threshold of $2d$ random percolation and for purely Gaussian distributed 
edge-weights, the geometric properties of the system spanning loops compare well to those of 
ordinary self-avoiding walks.
Lately, we performed further simulations for the NWP model on hypercubic lattice graphs in dimensions
$d\!=\!2$ through $7$ \cite{melchert2010a}, where we found evidence for an upper critical 
dimension $d_u\!=\!6$ of the NWP phenomenon.
To resume, up to now we have focused on the critical properties of the NWP model in the vicinity of the
critical point at which percolating loops first appear in the limit of large system sizes. 
As we experienced, at this critical point the loops are rather isolated and well separated 
from each other, i.e.\ they resemble a dilute gas of loops.
However, as the disorder on the lattice increases, the loops get more dense and it is reasonable 
to ask for the statistical properties of these densely packed configurations of loops. 

%
\begin{figure}[t!]
\includegraphics[width=1.0\linewidth]{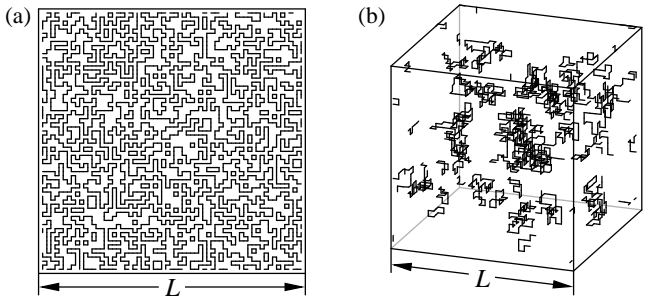}
\caption{Illustration of densely packed loop (DPL) configurations.
(a) Sample configuration of DPLs for a square lattice of side 
    length $L\!=\!64$ and fully periodic boundary conditions (BCs), and
(b) configuration of DPLs for a simple cubic lattice of side
    length $L\!=\!32$ and fully periodic BCs. For a more clear presentation, 
    small loops with length $\ell\!<\!10$ and a very large loop with length 
    $\ell\!=\!26\:770$ were omitted from the latter figure.}
\label{fig1ab}
\end{figure}  

Here, we investigate the NWP model on $2d$ square and honeycomb, $3d$ simple cubic and $4d$ hypercubic lattice graphs for 
a purely Gaussian distribution of the edge weights. Therein, our aim is to characterize the
minimum-weight configurations of densely packed loops (DPLs), see Fig.\ \ref{fig1ab}, regarding their 
statistical properties.
Further, we present a modification of the NWP algorithm that allows for the computation
of configurations of fully packed loops (FPLs).
Regarding a configuration of FPLs, each site on the lattice is visited by a loop.
We find that the scaling properties of DPLs and FPLs for the $2d$ systems agree within errorbars
and compare well to those obtained for FPLs related to other discrete interface models \cite{zeng1998}. 
Further, the geometric properties of the loops for $3d$ systems are similar to those of completely
uncorrelated random walks, indicating that the upper critical dimension shifts to $d_u^{\rm DPL}\!=\!3$
for DPLs.

The remainder of the presented article is organized as follows.
In section \ref{sect:model}, we introduce the model in more detail and 
we outline the algorithm used to compute the minimum weight configurations 
of loops (i.e.\ DPLs and FPLs). In section \ref{sect:results}, we present the results of 
our numerical simulations and in section \ref{sect:conclusions} we 
conclude with a summary.


\section{Model and Algorithm\label{sect:model}}

In the remainder of this article we consider regular $2d$ square/honeycomb,
$3d$ simple cubic and $4d$ hypercubic lattice graphs $G\!=\!(V,E)$ with side length $L$ and 
fully periodic boundary conditions (BCs).
The considered graphs have $N\!=\!|V| \!=\!L^d$ sites
$i\!\in\!V$ and a number of $|E| \!=\!z N/2$ undirected edges  
$\{i,j\}\!\in\!E$ that join adjacent sites $i,j\!\in\!V$.  
Above, $z$ signifies the coordination number of the lattice geometry, where
$z=4$, $3$, $6$ and $8$ for square, honeycomb, simple cubic and $4d$ hypercubic lattice graphs, 
respectively.
We further assign a weight $\omega_{ij}$ to each edge contained in $E$. 
These weights represent quenched random variables that introduce disorder 
to the lattice.
In the presented article we consider independent identically distributed 
weights drawn from a Gaussian disorder distribution with mean zero and unit width, i.e.\
\begin{equation}
P(\omega)=\exp{(-\omega^2/2)}/\sqrt{2\pi}. \label{eq:disorderDistrib}
\end{equation} 
Note that this distribution explicitly allows for loops $\mathcal{L}$ with a negative 
total weight $\omega_{\mathcal{L}}\!=\!\sum_{\{i,j\}\in\mathcal{L}}\omega_{ij}$.
So as to support intuition: Eq.\ \ref{eq:disorderDistrib} corresponds to the limit
$\rho\!=\!1$ of the disorder parameter for the edge-weight 
distribution considered in previous studies of the NWP model, 
see e.g.\ Ref.\ \cite{melchert2010a}. Accordingly, we here consider the NWP 
phenomenon high up in the percolating phase where the minimal weight configurations
of loops are densely packed.

\begin{figure}[t!]
\centerline{
\includegraphics[width=1.0\linewidth]{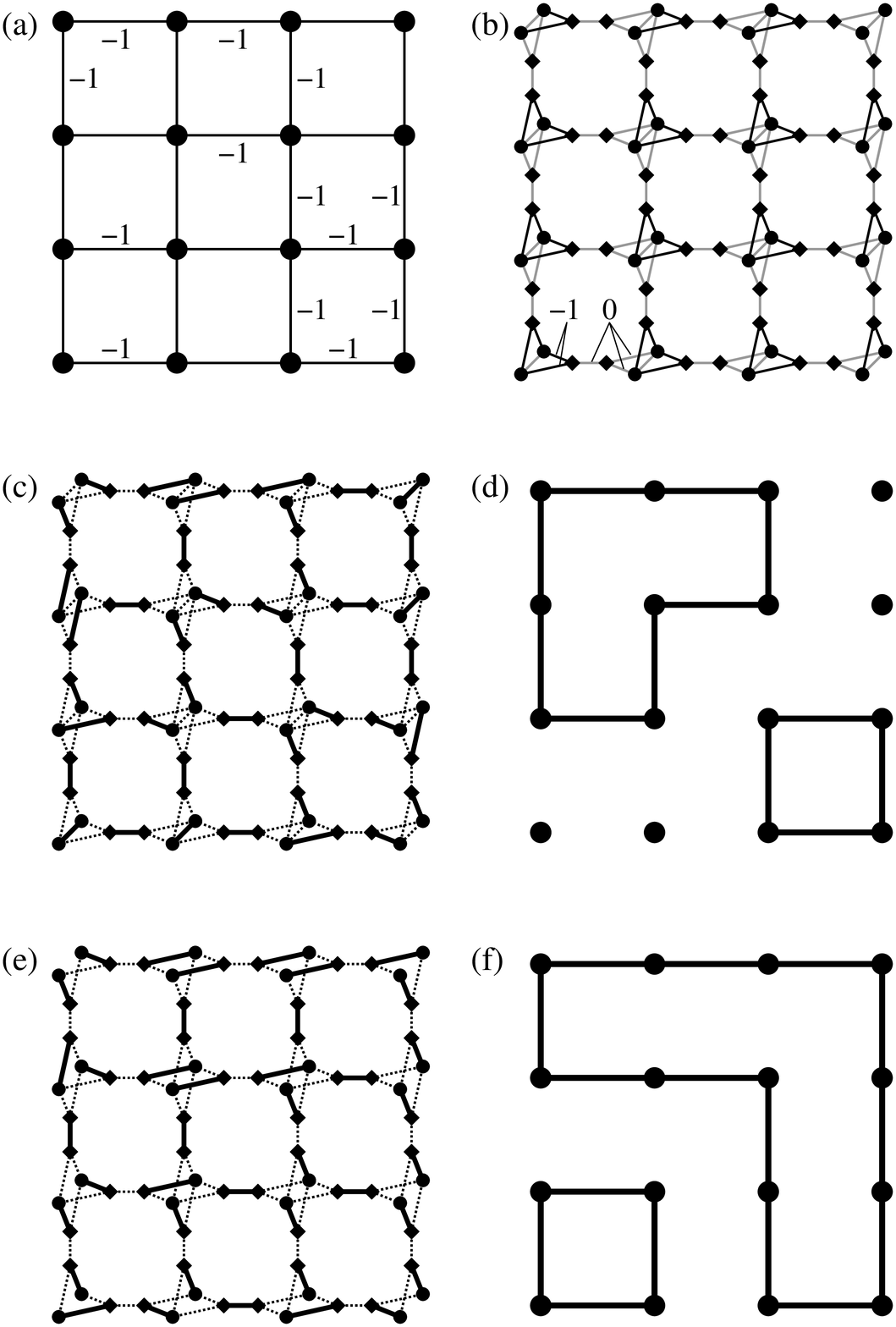}}
\caption{Illustration of the algorithmic procedure:
(a) original lattice $G$ with edge weights. 
For clarity, a bimodal distribution that yields edge-weights $\pm1$ is considered.
Further, only negative edge-weights are shown. Unlabeled edges have weight $+1$. 
(b) auxiliary graph $G_{\rm A}$ with proper weight assignment. Black 
edges carry the same weight as the respective edge in the original
graph and gray edges carry zero weight,
(c) minimum-weight perfect matching (MWPM) $M$: bold edges are matched 
and dashed edges are unmatched, and
(d) loop configuration (bold edges) that corresponds to the MWPM 
depicted in (c).
(e) MWPM for the the modified mapping that yields fully packed configurations
of loops.
(f) fully packed loops (bold edges) that correspond to the MWPM 
shown in (e).
\label{fig2abcd}}
\end{figure}

The NWP problem then reads as follows:
Given $G$ together with a realization of the disorder, determine a set 
$\mathcal{C}$ of loops such that the configurational energy, defined as the 
sum of all the loop-weights 
$\mathcal{E}\!=\!\sum_{\mathcal{L} \in \mathcal{C}} \omega_{\mathcal{L}}$, 
is minimized. As further optimization constraint, the loops are not
allowed to intersect.
Clearly, the configurational energy $\mathcal{E}$ is the quantity subject to
optimization and the result of the optimization procedure is a set of
loops $\mathcal{C}$, obtained using an appropriate transformation of
the original graph as detailed in \cite{ahuja1993}.  
For the transformed graphs, \emph{minimum-weight perfect matchings} (MWPMs)
\cite{cook1999,opt-phys2001,melchertThesis2009} are calculated, that 
serve to identify the loops for a given realization of the disorder. 
This procedure allows for an efficient implementation \cite{practicalGuide2009} of the 
simulation algorithms.
Until further notice, the subsequent description applies to DPLs only. 
In order to obtain FPLs, the transformation is slightly different (see discussion below).
Here, we give a brief description of the algorithmic procedure that yields a 
minimum-weight set of loops for a given realization of the disorder. 
Fig.\ \ref{fig2abcd} illustrates the 3 basic steps, detailed below:

(1) each edge, joining adjacent sites on the original graph $G$,  is
replaced by a path of 3 edges.  Therefore, 2  ``additional'' sites
have to be introduced for each edge in $E$.  Therein, one of
the two edges connecting an additional site to an original site gets
the same weight as the corresponding edge in $G$. The remaining  two
edges get zero weight.  The original sites $i\in V$ are then
``duplicated'',  i.e. $i \rightarrow i_{1}, i_{2}$, along with all
their incident edges and the corresponding weights. 
 For each of these pairs of duplicated sites,
one additional  edge $\{i_1,i_2\}$ with zero weight is added that
connects the two sites $i_1$ and $i_2$.  The resulting auxiliary graph
$G_{{\rm A}}=(V_{{\rm A}},E_{{\rm A}})$  is shown
in Fig.\ \ref{fig2abcd}(b), where additional sites appear as squares and
duplicated  sites as circles. Fig.\ \ref{fig2abcd}(b) also illustrates
the weight  assignment on the transformed graph $G_{{\rm A}}$.  Note
that while the original graph (Fig.\ \ref{fig2abcd}(a)) is symmetric, the
transformed graph (Fig.\ \ref{fig2abcd}(b)) is not. This is due to the
details of the mapping procedure and the particular weight assignment
we have chosen.  A more extensive description of the mapping can be
found in \cite{melchert2007}.

(2) a MWPM on the auxiliary graph is determined via exact
combinatorial optimization algorithms \cite{comment_cookrohe}.  A MWPM
is a minimum-weighted subset $M$ of $E_{\rm A}$, such that
each site  contained in $V_{\rm A}$ is met by precisely one
edge in $M$.  This is illustrated in Fig.\ \ref{fig2abcd}(c), where the
solid edges  represent $M$ for the given weight assignment. The dashed
edges are  not matched.  Due to construction, the auxiliary graph
consists of an even number of sites and 
the transformation procedure described in step (1) guarantees
that a perfect matching exists.  
Note that a MWPM can be computed in polynomial time as a function
of the number of sites, hence large systems with several thousands
of sites are feasible.

(3) finally it is possible to find a relation between the matched
edges $M$  on $G_{\rm A}$ and a configuration of negative-weighted
loops  $\mathcal{C}$ on $G$ by  tracing back the steps of the
transformation (1). As regards this, note that each edge  contained
in $M$ that connects an additional site (square) to a duplicated  site
(circle) corresponds to an edge on $G$ that is part of a loop, see
Fig.\ \ref{fig2abcd}(d).
Note that, by construction of the auxiliary graph,
 for each site $i_1$ or $i_2$ matched in this
way, the corresponding twin site $i_2$/$i_1$ must be matched
to an additional site as well. This guarantees that wherever a path enters
a site of the original graph, the paths also leaves the site, corresponding
to the defining condition of loops.
All the edges in $M$ that connect like  sites (i.e.\ duplicated-duplicated, or
additional-additional)  carry zero weight and do not contribute to a
loop on $G$.  
Once all loop segments are found, a depth-first search \cite{ahuja1993,opt-phys2001} can  be used to
identify the loop set $\mathcal{C}$ and to determine the geometric 
properties of the individual loops. 
Here, the exemplary weight assignment illustrated 
in Fig.\ \ref{fig2abcd}(a) yields 2 loops, 
i.e.\ $\mathcal{C}\!=\!\{\mathcal{L}_1, \mathcal{L}_2\}$, with
weights $\omega_{\mathcal{L}_1}\!=\!\omega_{\mathcal{L}_2}\!=\!-4$
and lengths $\ell_1\!=\!\sum_{\{i,j\}\in\mathcal{L}_1}1\!=\!8$, 
$\ell_2\!=\!4$. Hence, the configurational energy reads $\mathcal{E}\!=\!-8$. 

The result of the calculation is a collection $\mathcal{C}$
of loops such that the total loop weight, and consequently the
configurational energy $\mathcal{E}$, is minimized. 
Hence, one obtains a global collective optimum of the system.  
Obviously, all loops that contribute to $\mathcal{C}$ possess a negative weight.  
Also note that the choice of the weight assignment in step (1) is not unique, 
i.e.\ there are different possibilities to choose a weight assignment
that all result in equivalent sets of matched edges on the transformed
lattice, corresponding to the minimum-weight collection of loops on
the original lattice. Some of these weight assignments result in a more
symmetric  transformed graph, see e.g. \cite{ahuja1993}. However, this
is only a technical issue that does not affect the resulting loop
configuration. Finally, for an illustrational purpose, a small $2d$ lattice graph with
free BCs was chosen intentionally. The algorithmic procedure extends 
to higher dimensions and fully periodic BCs in a straight-forward manner.

Note that, by means of the algorithmic procedure outlined above, its also 
possible to compute configurations of fully packed loops. 
To accomplish this, only one detail related to step (1) has to be altered.
I.e., original sites $i\!\in\!V$ are duplicated ($i\!\to\!i_1,i_2$) to 
yield $i_1,i_2\!\in\!V_A$, but \emph{no} additional edge $\{i_1,i_2\}$ is added to $E_A$.
By construction, a MWPM, as computed in step (2), will not contain an edge
that connects a tuple of duplicated sites (since there are no such edges present). Hence, in the resulting MWPM,
each duplicated site will be matched to an additional site, see Fig.\ \ref{fig2abcd}(e). 
Applying step (3), one then finds that a each site on the original graph 
is an endpoint of exactly two loop segments, see Fig.\ \ref{fig2abcd}(f), 
and the result is a fully packed configuration of loops.
Applying this modified mapping to the exemplary weight assignment illustrated 
in Fig.\ \ref{fig2abcd}(a) yields $\mathcal{C}_{\rm FPL}\!=\!\{\mathcal{L}_1, \mathcal{L}_2\}$ (see Fig.\ \ref{fig2abcd}(f)), 
with weights $\omega_{\mathcal{L}_1}\!=\!-6$, $\omega_{\mathcal{L}_2}\!=\!0$
and lengths $\ell_1\!=\!12$, $\ell_2\!=\!4$. Here, the configurational energy reads $\mathcal{E}\!=\!-6$. 

Apparently, there are qualitative differences between configurations of 
DPLs and FPLs: Regarding the DPLs, not all lattice sites are visited by a loop 
(see Fig.\ \ref{fig1ab}(a)), and, due to the ``energy minimization principle'' of 
the optimization procedure, the weight of an individual loop is strictly smaller than zero. 
In contrast, FPLs require each lattice site to be visited by a loop. This constitutes 
a ``hard'' constraint for the optimization procedure and as a result, an individual 
loop weight might take a value larger than or equal to zero (as for the assignment of $\pm 1$ 
edgeweights in Fig.\ \ref{fig2abcd}(a)).

Subsequently, we will use the procedure outlined above so as to 
investigate the NWP model on $2d$ square and honeycomb and $3d$ simple cubic lattice
graphs with a Gaussian distribution of edge weights. 

\section{Results \label{sect:results}}

In order to characterize the statistical properties of configurations of DPLs
we performed simulations for $2d$ square ($2d$-sq) systems with a linear extension of up 
to $L\!=\!256$ lattice sites, and $3d$ simple cubic ($3d$-sc) systems with side length 
up to $L\!=\!48$ sites. 
If not stated explicitly, the results presented in the remainder of this section 
follow from an analysis of the full ensemble of loops obtained for a fixed system size, 
i.e.\ $L\!=\!256$ ($48$) in $2d$ ($3d$). For both setups we considered a number 
of $n\!=\!12800$ independent realizations of the disorder to compute averages.
Moreover, we conducted simulations for a $2d$ honeycomb ($2d$-hy) DPL and $2d$-sq FPL version 
of the loop model for lattices up to $L\!=\!128$, where we considered $n\!=\!9600$
samples for the largest systems.
We performed additional simulations for $4d$ hypercubic ($4d$-hc) lattice graphs with $L\!=\!20$ and
$n\!=\!9600$ in order to study selected observables for an even higher dimension.

Next, we briefly introduce the basic observables we used to characterize the 
individual loops obtained from the simulations. As introduced earlier, 
two fundamental quantities related to a given loop are its weight $\omega_{\mathcal{L}}$ 
and length $\ell$.
We further determine the linear extensions $R_i$, $i\!=\!1\ldots d$, 
of each loop by projecting it onto the independent lattice axes 
(for the hexagonal lattice geometry we therefore use the topologically
equivalent brickwall lattice).
The largest of those values, i.e.\ 
\begin{align}
R\!=\!\max_{i=1\ldots d}(R_i), 
\end{align}
is referred to as the \emph{spanning length} of the loop and the smallest, 
i.e.\ 
\begin{align}
r\!=\!\min_{i=1\ldots d}(R_i), 
\end{align}
is called its \emph{roughness}. 
Moreover, so as to characterize the full perimeter of an individual loop 
on a coarse grained scale, we can define the \emph{size} 
\begin{align}
R_s\!=\!\sum_{i=1}^{d} R_i,  
\end{align}
i.e.\ the length of the loop if all small scale irregularities were flattened \cite{vachaspati1984}.
Note that the linear extensions of a given loop are measured in units 
of lattice sites. E.g., the loop $\mathcal{L}_1$ contained in the 
example illustrated in Fig.\ \ref{fig2abcd}(d) has $R_1\!=\!R_2\!=\!3$ and 
thus $R_s\!=\!6$. Albeit this definition of a loop size is somewhat 
odd for small loops in a $3d$ setup (where an elementary loop of
length $\ell\!=\!4$ has size $R_s\!=\!5$, as noted in Ref.\ \cite{vachaspati1984}), 
it nevertheless gives a reasonable definition of a coarse grained loop size 
in the limit of large $\ell$.
Anyway, note that there are some difficulties related to the ``extremal'' loops
on the lattice. This means, if we consider the full ensemble of loops for a system
of size $L$, there are very small loops ($R\!\ll\!L$) which are affected
by the finite lattice spacing and also very large loops ($R\!\approx\!L$) 
that are affected by the finite size of the considered system. So as to obtain 
most reliable estimates for the observables that characterize the configurations
of loops, and if not stated explicitly, these extremal loops are discarded from the analysis.
Subsequently, we will adopt the notation $\langle O \rangle$ to signify the disorder averaged
value of observable $O$. Whenever a more clear presentation requires, we will
write $\langle O \rangle_X$ to signify the disorder averaged value of $O$ with 
respect to the quantity $X$.

In the following we will present the results on the geometric properties of the 
DPL/FPL configurations in subsection \ref{subsec:Results_geom}, the results on 
the energetic properties are summarized in subsection \ref{subsec:Results_erg}.

\subsection{Geometric properties}\label{subsec:Results_geom}

As pointed out above, we here report our results regarding the geometric properties of the
loop configurations obtained for the $2d$ ($3d$, $4d$) systems mention in the beginning
of the presented section.

%
\begin{table*}
\caption{
``Packing'' properties, scaling exponents and parameters that characterize DPLs/FPLs
on the considered lattice graphs. From left to right:
System setup [sq=square, hy=honeycomb, sc=simple cubic, hc=hypercubic;
DPLs (FPLs)=densely (fully) packed loops], 
linear extension $L$ of the lattice for which the listed 
estimates were obtained, number $n$ of samples considered,
approximate number $N_{\rm loops}$ of loops collected during the simulations,
probability $\rho_{\rm cov}$ that a lattice site is visited 
by a loop, fraction $f_\infty$ of loop segments contained in 
system spanning loops and ``loop-shape'' parameter $a/(2d^2)$.
Fractal dimension $d_f$, correction to scaling exponent $\omega$ 
and exponent $\tau$ that describe the scaling of the non-spanning loops.
Scaling dimension $d_f^\prime$ for the largest
loops found for each realization of the disorder and asymptotic 
estimate $\epsilon$ for the normalized configurational weight.
} 
\label{tab:tab2}
\begin{tabular}[c]{lllllllllllll}
\hline\noalign{\smallskip}
      	& & $L$ & $n$ & $N_{\rm loops}$ & $\rho_{\rm cov}$ & $f_\infty$ & $a/(2d^2)$ & $d_f$ & $\omega$ & $\tau$ & $d_f^\prime$ & $\langle \epsilon \rangle$ \\
\noalign{\smallskip}\hline\noalign{\smallskip}
$2d$-sq &DPLs & $256$ 		& $12\:800$ 		& $25\times10^6$    	& $0.825$ & $0.327$ & $0.97(3)$ & 1.756(8) & 0.86(8)  & 2.14(2) & 1.752(1) & 0.596 \\
$2d$-sq &FPLs & $128$ 		& \phantom{0}$9\:600$ 	& \phantom{0}$6\times10^6$ 	& $1.000$ & $0.371$ & $0.98(1)$ & 1.75(2)  & 0.9(2)   & 2.16(1) & 1.751(1) & 0.493 \\
$2d$-hy &DPLs & $128$ 		& \phantom{0}$9\:600$ 	& \phantom{0}$2\times10^6$ 	& $0.750$ & $0.429$ & $1.04(4)$  & 1.75(2)  & 1.0(1)   & 2.14(1) & 1.750(1) & 0.412 \\
$3d$-sc &DPLs & \phantom{0}$48$ & $12\:800$ 		& $13\times10^6$ 		& $0.901$ & $0.907$ & $0.96(1)$ & 2.01(4)  & 0.66(11) & 2.49(6) & 3.00(1)  & 0.844 \\
$4d$-hc &DPLs & \phantom{0}$20$ &  \phantom{0}$9\:600$ 	& \phantom{0}$6\times10^6$ 	& $0.938$ & $0.970$ & $0.99(3)$ & 2.00(5)  & -        & -       & -        & 1.013     \\
\noalign{\smallskip}\hline
\end{tabular}
\end{table*}
\paragraph{Global ``packing'' properties:} 
As pointed out above, for the configurations of DPLs considered here,
not all lattice sites are visited by a loop.  We can quantify this by
computing the probability $\rho_{\rm cov}$ that some site on the
lattice lies on the perimeter of a loop, see Tab.\
\ref{tab:tab2}. 
Via additional simulations for other system sizes (not shown here), 
we verified that
the value of $\rho_{\rm cov}$ is basically independent of the lattice
sizes.  From the accumulated data we also obtained the
probability $\rho_\infty$ that some site on the lattice lies on the
perimeter of a large, i.e.\ system spanning, loop.  From this we
conclude that a fraction $f_\infty\!\equiv\!\rho_\infty/\rho_{\rm
cov}$ (see Tab.\ \ref{tab:tab2})  of all loop segments for the
$2d$-sq, $2d$-hy, $3d$-sc and $4d$-hc setup is  contained in large
loops.  Further, note that the estimate $f_\infty\!\approx\!0.371$ for
the $2d$-sq FPL version of the loop model  (where $\rho_{\rm
cov}\!=\!1$, exactly) is somewhat larger than  the corresponding value
$f_\infty\!\approx\!0.327$ for the DPL configurations.  As evident by
comparing the estimates $\rho_{\rm cov}$ and $f_\infty$ obtained for
DPLs  on $2d$ square and honeycomb lattice graphs, the observables
above characterize the particular  lattice geometries chosen and hence
are not universal.

\paragraph{Scaling of the loop shape:} 
Next, we want to study the scaling behavior of the average loop shape. 
As in Refs.\ \cite{vachaspati1984,melchert2010a} we therefore monitor the volume to 
surface ratio $V_{\rm B}/S_{\rm B}$ of the smallest box that fits the individual 
loops as a function of the coarse-grained loop size $R_s$, where 
$V_{\rm B}\!=\!\Pi_{i=1}^{d}R_i$ and $S_{\rm B}\!=\!2\times\sum_{i=1}^{d} V_{\rm B}/R_i$.
For hypercubic volumes with identical values $R_1\!=\!\ldots\!=\!R_d$, one would expect to find 
$V_{\rm B}/S_{\rm B}\!=\!(2d^2)^{-1}R_s$. 
Performing fits to the form $\langle V_{\rm B}/S_{\rm B}\rangle\!=\!a R_s^{\psi}$, discarding 
very small and large nonpercolating loops, we here yield the results
$\psi\!=\!1.002(5)$ ($1.001(1)$) and $a\!=\!0.121(3)$ ($0.0535(3)$) that 
describe $2d$-sq ($3d$-sc) configurations of DPLs. 
Further, for the $2d$-sq FPL data we obtain $\psi\!=\!0.999(2)$ and $a\!=\!0.123(1)$.
For the $2d$-hy setup our data is consistent with the precise scaling expression 
$\langle V_{\rm B}/S_{\rm B}\rangle\!=\! 0.130(5) (R_s-0.4)^{1.006(8)}$, that 
includes corrections to scaling in a very basic form. The same holds also for the 
$4d$-hc setup, where $\langle V_{\rm B}/S_{\rm B}\rangle\!=\! 0.031(1) (R_s-0.5)^{0.994(8)}$.
Hence, we find values of $a/(2d^2)$ (see Tab.\ \ref{tab:tab2}) reasonably close to $1$ in order 
to conclude that, in a statistical sense, the loops are not oblate but possess 
a rather spherical shape.

\begin{figure}[t!]
\centerline{
\includegraphics[width=1.0\linewidth]{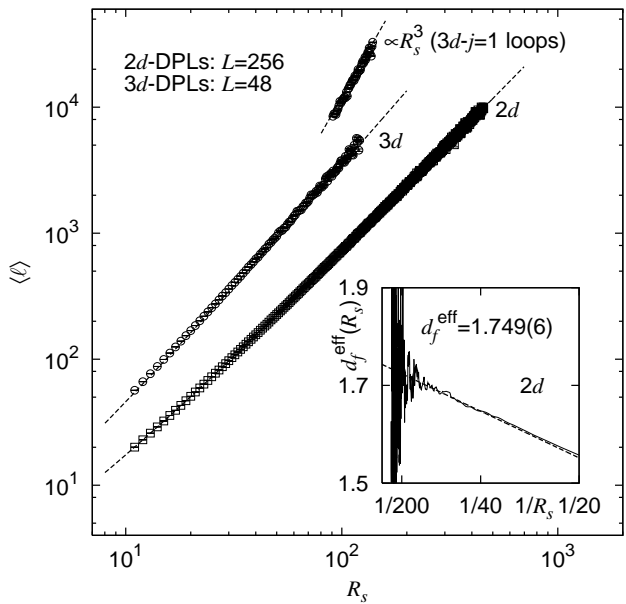}}
\caption{ \label{fig:fig3}
Scaling of the average loop length $\langle \ell \rangle$ of the non-spanning DPLs
as a function of the coarse grained loop size $R_s$ for $2d$ square and 
$3d$ simple cubic lattice graphs (the $3d$ data is shifted upwards by a 
factor of $3$). Further, the upmost data curve indicates the scaling $\sim\!R_s^3$ 
of $3d$ loops with spanning index $j\!=\!1$, i.e.\ those loops that span the 
system along exactly one direction (the respective data points are shifted upwards 
by a factor of $12$).
The inset illustrates the sequence of effective scaling exponents $d_f^{\rm eff}$
as function of the inverse loop size for the DPL setup on the $2d$ square lattice.}
\end{figure}

\paragraph{Scaling of the loop length -- ``small'' loops:} 
The geometric properties of the loops can further be characterized by the  
fractal dimension $d_f$. It can be defined from the scaling of the average 
loop length $\langle\ell\rangle$ as a function of the loop size according to 
$d_f\!=\! \lim_{R_s \to \infty} \log(\langle \ell \rangle)/\log(R_s)$. 
Lower and upper bounds on the above estimate of $d_f$ can be obtained in a systematic
way \cite{camarda2006}: Based on the expression $R_s=\sum_{i=1}^d R_i$ for the loop size,
it is intuitive that $R_s^-=d\cdot r$ typically underestimates the value of $R_s$ (bear in mind that $r=\min_{i=1\ldots d}(R_i)$). 
Consequently, the scaling assumption $\langle \ell \rangle_r \sim r^{d_f^-}$ yields an estimate 
$d_f^-\leq d_f$. In the same sense, $R_s^+=d\cdot R$ overestimates $R_s$ and gives
rise to a scaling exponent $d_f^+\geq d_f$. However, 
in the limit of large loops and since the loops have rather
spherical shape (which implies $r\approx R$), we should be able to verify $d_f^-=d_f=d_f^+$.
For the $2d$ and $3d$ systems considered, we found our data best fit by a scaling function
that includes corrections to scaling in the form $\langle\ell\rangle \sim X^{d_f}(1+cX^{-\omega})$, 
where $X$ optionally stands for $r$, $R_s$ or $R$. 
In this regard we yield the estimates 
listed in Tab.\ \ref{tab:tab3}
and fit-parameters $c$ of the order of unity 
(Note that for the $4d$ hypercube with $L=20$, since the 
range of values for the scaling parameters $r$ and $R$ is too small, 
our data did not allow for a decent analysis of upper and lower bounds). 
The exponents for the choice $X=R_s$ are listed in Tab.\ \ref{tab:tab2} and
the data for the $2d$-sq ($3d$-sc) DPLs is shown in Fig.\ \ref{fig:fig3}. 
When comparing the resulting estimates of $d_f^{+/-}$ and $d_f$, keep
in mind that the presence of corrections to scaling hinders the straight forward 
fit to a pure power law function.
%
\begin{table}[b!]
\caption{\label{tab:tab3}
Extended analysis of scaling exponents and parameters that characterize DPLs/FPLs.
From left to right: 
lattice setup [sq=square, hy=honeycomb, sc=simple cubic; DPLs (FPLs) = densely (fully) packed loops], 
lower and upper bounds $d_f^{-/+}$ on the fractal dimension and corresponding correction to scaling 
exponents $\omega^{+/-}$ that describe the scaling of the non-spanning loops as explained in the text.
} 
\begin{tabular}[c]{llllll}
\hline\noalign{\smallskip}
 &  & $d_f^-$ & $\omega^-$ & $d_f^+$ & $\omega^+$  \\
\noalign{\smallskip}\hline\noalign{\smallskip}
2d-sq &DPLs & 1.749(6)  & 0.8(3)  & 1.751(5)  & 0.92(5)\\
2d-sq &FPLs & 1.73(2)   & 1.1(3)  & 1.75(1)   & 0.7(1) \\
2d-hy &DPLs & 1.73(4)   & 1.0(3)  & 1.75(2)   & 1.1(1) \\
3d-sc &DPLs & 2.0(1)    & 0.5(2)  & 2.0(3)    & 0.3(4)  \\
\noalign{\smallskip}\hline
\end{tabular}
\end{table}
As an alternative to the scaling function considered above, one might choose a more basic 
expression that accounts for corrections to scaling and features less free parameters.
In this regard, for the $2d$-sq FPL setup we obtained 
$\langle\ell\rangle \sim (R_s+1.5(4))^{1.747(9)}$, in agreement with the estimate listed in Tab.\ \ref{tab:tab2}.
From the simulations for the $4d$ systems and in the limit of large loops, 
we found a good fit of the data to the form $\langle \ell \rangle \sim R_s^{d_f}$, 
resulting in the estimate $d_f\!=\!2.00(5)$.

As further check of the results for the $2d$-sq ($3d$-sc) DPLs obtained above, we computed the effective (local) slopes 
that describe the scaling of $\langle \ell \rangle$ within intervals 
of, say, $m$ consecutive data points $R_s$ according to $\log(\langle \ell \rangle)\!=\! d_f^{\rm eff} \log(R_s) + c$.
By sliding this ``scaling window'' over the whole data set, one obtains a sequence of effective
exponents $d_f^{\rm eff}(R_s)$. The asymptotic value $d_f^{\rm eff}$ is then extrapolated 
by means of a straight line fit to the plot of $d_f^{\rm eff}(R_s)$ as a function of the 
inverse loop size. For the $2d$-sq ($3d$-sc) DPL data, setting $m\!=\!20$ ($10$), this procedure yields 
the estimates $d_f^{\rm eff}\!=\!1.749(6)$ ($1.98(3)$), that agree with those listed
in Tab.\ \ref{tab:tab2} within errorbars. The results for the DPLs on the $2d$ square lattice 
are further shown in the inset of Fig.\ \ref{fig:fig3}.
\begin{figure}[t!]
\centerline{
\includegraphics[width=1.0\linewidth]{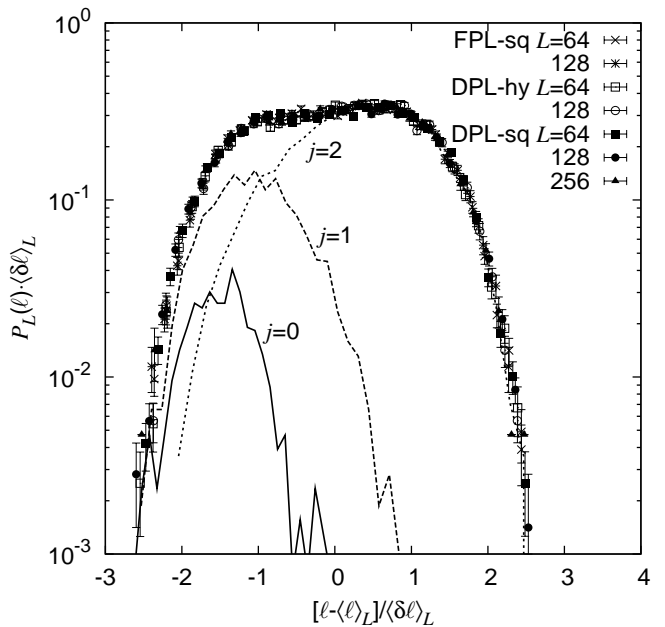}}
\caption{ \label{fig:fig6}
Scaling behavior of the probability density function $P_L(\ell)$ of the loop length $\ell$,
considering the largest loop found for each realization of the disorder for
different system sizes $L$. After rescaling, data curves for
the different $2d$ setups 
[DPLs (FPLs) = densely (fully) packed loops; sq=square, hy=honeycomb] 
fall onto one universal scaling
function. Further, the contribution of the loops with spanning
index $j\!=\!0,1,2$ (see text) to the $L\!=\!256$ ($2d$ DPL-sq) data curve is illustrated.
}
\end{figure}  

Referring to the numerical values of $d_f$ found here, and 
from a point of view of statistical physics, the loops appear to be self similar 
fractals.
Albeit the loops are self avoiding by construction, our results indicate that for $d\!\geq\!3$ 
they exhibit the statistical properties of completely uncorrelated random curves with $d_f\!=\!2$.
Hence, this suggests that the upper critical dimension of the NWP problem in the limit of 
DPLs is $d_u^{\rm DPL}\!=\!3$.

\paragraph{Scaling of the loop length -- ``large'' loops:} 
So far, the data analyses referred to the subset of non-spanning loops found for the simulated
disorder samples.
However, as discussed above, about $91\%$ of the over all loop length for the $3d$-sc 
systems is contained in very large, i.e.\ system spanning, loops (see Tab.\ \ref{tab:tab2}).
For this setup we further observe that, in comparison to the small loops investigated above, 
very large system spanning loops exhibit different scaling properties.
These loops are a direct consequence of the periodic boundary conditions, and, 
although they are strongly affected by the finite size of the ``simulation box'',
we might nevertheless characterize their configurational properties. 

To perform a systematic analysis of these very large loops, we here introduce the
spanning index $j\!\in\!\{0\ldots d\}$ of an individual loop. As regards this, a value of, say, 
$j\!=\!2$ indicates that the respective loop spans the lattice along exactly $2$ of the $d$
independent lattice directions (non-spanning loops have $j\!=\!0$). 
Note that for a single configuration of the disorder, the number of loops having $j\!\geq\!1$ 
is rather small. E.g., simulating $n\!=\!12\:800$ instances
for a $3d$ simple cubic system with $L\!=\!48$ we count $N_{\rm loops}^j\!=\!4\:149$, $3\:260$, $31\:106$
for $j\!=\!1$, $2$, $3$, respectively (for comparison $N_{\rm loops}^{j=0}\!=\!13\:665\:016$).
Hence, the results below should be recognized as ``guiding values'' that are to be
encountered with some caution. 

Now, filtering the $3d$ data for loops with spanning index $j\!=\!1$, we observe a power law 
scaling of the form $\langle\ell\rangle\!\sim\!R_s^{3.0(1)}$, see Fig.\ \ref{fig:fig3}.
Considering a loop with spanning index $j\!=\!2$, one might argue that the roughness $r$ is
a more adequate scaling variable. Note that for loops with $j\!=\!2$ on a lattice of side length 
$L\!=\!48$ one has $r=R_s-2\cdot48$.
As regards this, upon analysis we find a best fit of 
the data to the form $\langle \ell\rangle\!\sim\! (R_s-91(8))^{1.7(3)}$. 
However, as mentioned above, the number of loops having $j\!=\!2$ is rather small. 
Hence the statisics for these loops is rather poor and the above estimate presents
only an approximate result.
Note that no such scaling analysis is possible for a spanning index $j\!=\!d$, since
the corresponding loops all have $R_s\!=\!d\cdot L$.

Regarding the $4d$ data we yield
$N_{\rm loops}^{j}\!=\!6\:625\:020$, $2\:873$, $1\:939$, $2\:021$ and $30\:961$
for $j\!=\!0,\ldots,4 $, respectively. 
Due to the small number of loops having $j\!=\!1,2,3$, no decent scaling analysis 
was possible.

We also performed FSS analyses according to $\langle \ell
\rangle\!\sim\! L^{d_f^\prime}$, where, for different system sizes
$L$, we consider only the largest loop found for each realization of
the disorder.  In this regard we were able to verify the ``small loop''
scaling behavior for the $2d$ systems,  i.e.\
$d_f^\prime\!\approx\!1.75$, see Tab.\ \ref{tab:tab2}. 
A similar scaling analysis for the $3d$-sc data
yields $\langle \ell \rangle\!\sim\!L^{3.00(1)}$ (not shown), where
all loops considered 
(i.e.\ the largest loop for each realization of the disorder)
turned out to have $j\!=\!3$. 
In $4d$ we performed simulations for $L\!=\!20$ only. Hence no 
such scaling analysis is carried out for the $4d$ setup.

 The probability density function (pdf) $P_L(\ell)$ of the length $\ell$
for the largest loop for a given realization of the disorder 
is shown in Fig.\ \ref{fig:fig6}. As evident from the figure,
the data for all the $2d$  setups considered fall onto one master
curve, if rescaled according to
$P_L(\ell)\!=\!\langle\delta\ell\rangle_L^{-1}p_0[(\ell-\langle\ell\rangle_L)/\langle\delta\ell\rangle_L]$.
Therein, $p_0[\cdot]$ signifies a scaling function and $\langle
\delta\ell\rangle_L\!=\![\langle\ell^2\rangle_L-\langle\ell\rangle^2_L]^{1/2}$
refers to the root mean square deviation related to the loop length
$\ell$.  This data collapse highlights that the different
geometries and types of loops exhibit an universal behavior.
We further observed that each data curve is composed of three
parts that relate to loops with different values of $j$.  For the $2d$
DPLs on the $L\!=\!256$ square lattice, Fig.\ \ref{fig:fig6} also
shows the contribution stemming from loops with spanning indices
$j\!=\!0$, $1$ and $2$. Loops with $j=0$ contribute to the shortest
loops, $j=2$ the largest loops and $j=1$/$j=2$ contribute to
the bulk of itermediate size loops.

\begin{figure}[t!]
\centerline{
\includegraphics[width=1.0\linewidth]{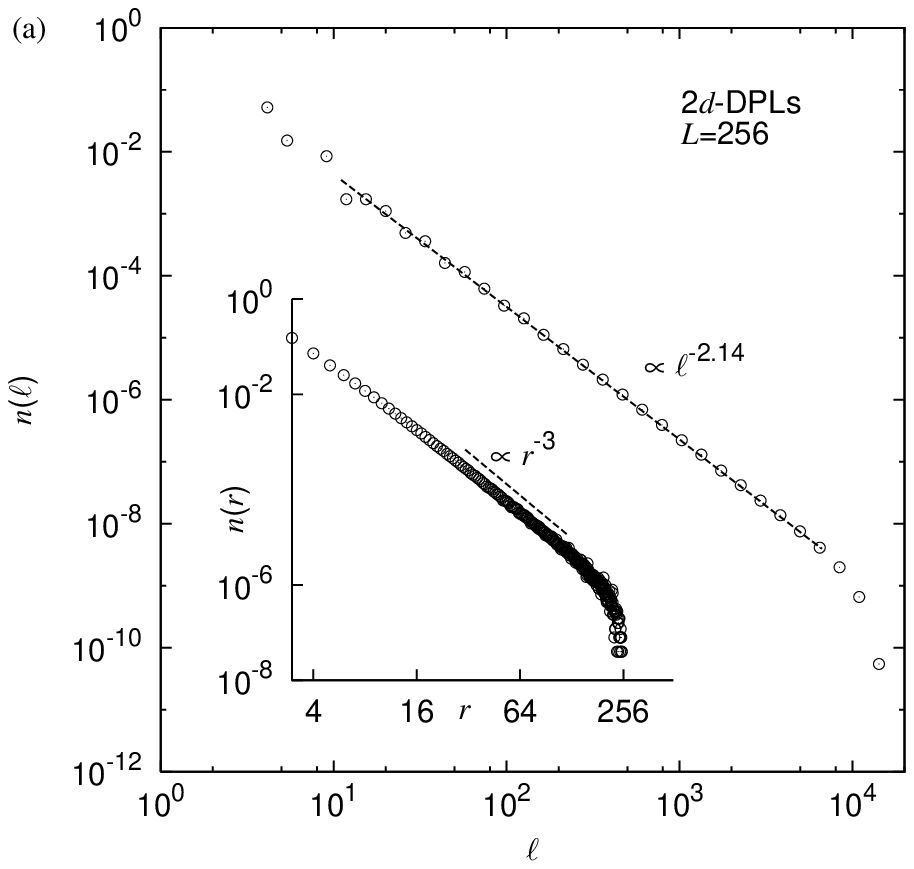}}
\centerline{
\includegraphics[width=1.0\linewidth]{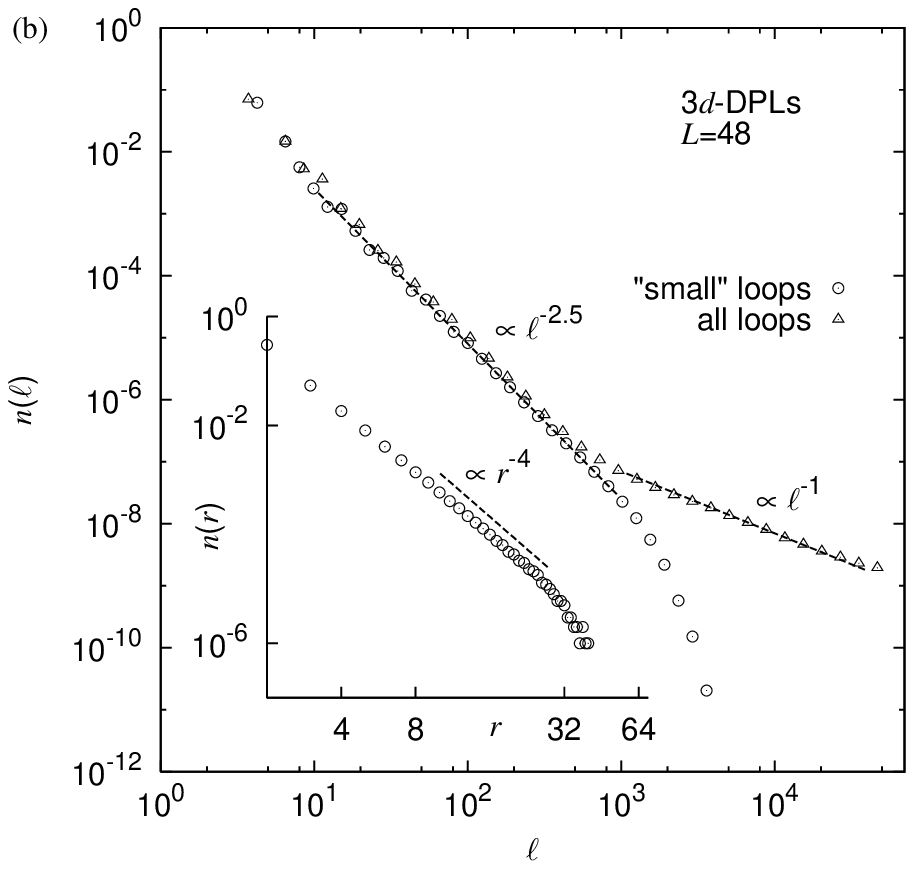}}
\caption{ \label{fig:fig4}
Results obtained for the number densities $n(\cdot)$ for 
loops of length $\ell$ and roughness $r$.
The main plots show the number density $n(\ell)$ of loops with 
length $\ell$ (using logartihmic binning) and the inset illustrates the scaling of the 
number density $n(r)$ of loops with roughness $r$.
(a) Data obtained for DPLs on the $2d$ square lattice, 
(b) data for DPLs on the $3d$ simple cubic lattice.}
\end{figure}  

\paragraph{Scaling of the loop area in $2d$:}
For the $2d$ setups we can further associate a \emph{cluster} with each loop. 
Such a cluster simply consists of the elementary plaquettes enclosed by 
a loop (for a square lattice, an elementary plaquette consists of a unit 
square bordered by 4 lattice edges).
Consequently, the loop forms the perimeter of its associated cluster. 
The basic observable related to a cluster is its area (or similarly its volume) $A_{\mathcal{L}}$, measured by
the number of elementary plaquettes that comprise the cluster.
To support intuition, the exemplary loop configuration $\mathcal{C}\!=\!\{\mathcal{L}_1,\mathcal{L}_2\}$
shown in Fig.\ \ref{fig2abcd}(d) features two clusters with perimeter $\ell_1\!=\!8$, $\ell_2\!=\!4$
and area $A_{\mathcal{L}_1}\!=\!3$, $A_{\mathcal{L}_2}\!=\!1$.

As introduced earlier, the spanning length $R$ is a typical lengthscale that characterizes
an individual loop, and henceforth characterizes also the associated cluster. The area of the clusters is
expected to scale as $A_{\mathcal{L}}\!\sim\!R^{d_v}$, where $d_v$ signifies the 
fractal dimension of the cluster area (or volume).
For $2d$-sq DPLs we find $\langle A_{\mathcal{L}}\rangle\!\sim\!R^{1.997(4)}$
($2d$-hy DPLs: $d_v\!=\!2.01(1)$; $2d$-sq FPLs: $d_v\!=\!1.99(1)$),
indicating that the clusters are compact with $d_v\!=\!d$.
Regarding the fractal scaling of the cluster perimeter ($\ell\!\sim\!R^{d_f}$),
we might further write $A_{\mathcal{L}}\!\sim\!\ell^{d/d_f}$, see Refs.\ \cite{leibler1987,duplantier1990}.
More precise, to account for corrections to scaling we consider a scaling function 
$\langle A_{\mathcal{L}}\rangle\!=\!A_0(\ell+\Delta\ell)^{\kappa}$ and obtain
\begin{align*}
& & A_0 &\; & \Delta\ell\ &\; & \kappa\ & & \\
{\rm DPLs-sq}&:	&0.378(2)&\; &-1.58(6)&\; & 1.137(1)\\
{\rm FPLs-sq}&:	&0.323(2)&\; &-0.71(4)&\; & 1.135(1)\\
{\rm DPLs-hy}&:	&0.397(9)&\; &-2.2(2) &\; & 1.135(4)
\end{align*}
Here, since $d_f\!=\!d/\kappa$, the scaling exponent $\kappa$ provides an independent 
way to estimate the fractal dimension of the loops.
In any case, the numerical value of $d/\kappa$ is in good agreement with the previous 
estimate of $d_f$ for the respective setup listed in Tab.\ \ref{tab:tab2}. 

\begin{figure}[t!]
\centerline{
\includegraphics[width=1.0\linewidth]{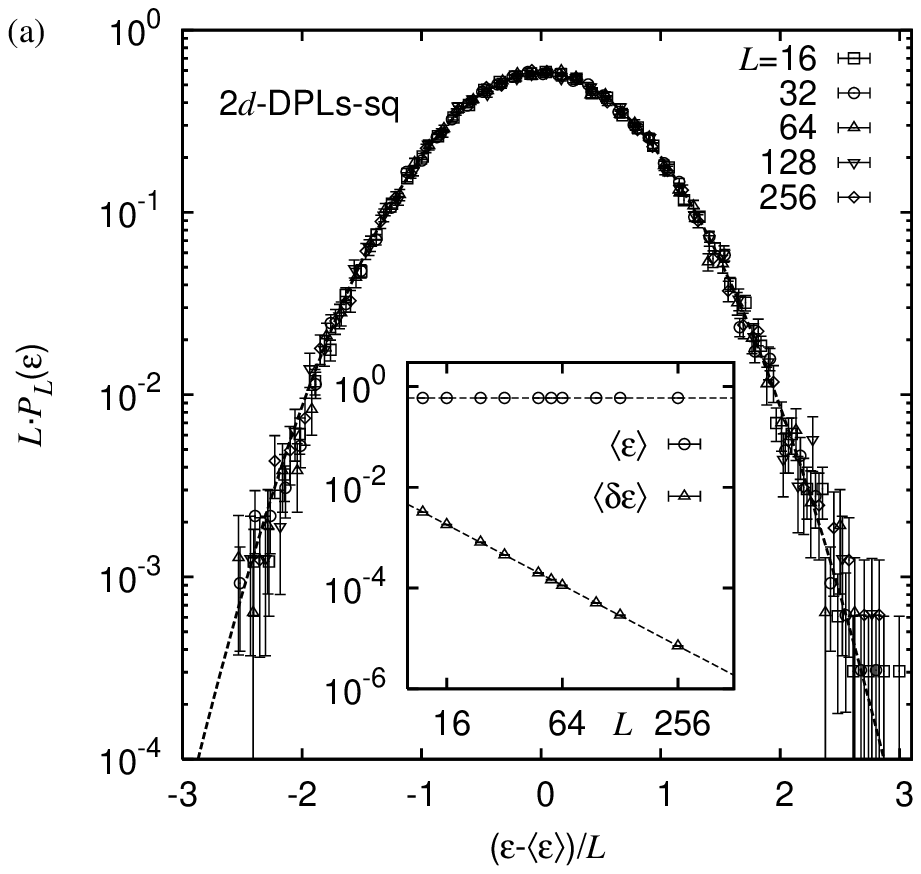}}
\centerline{
\includegraphics[width=1.0\linewidth]{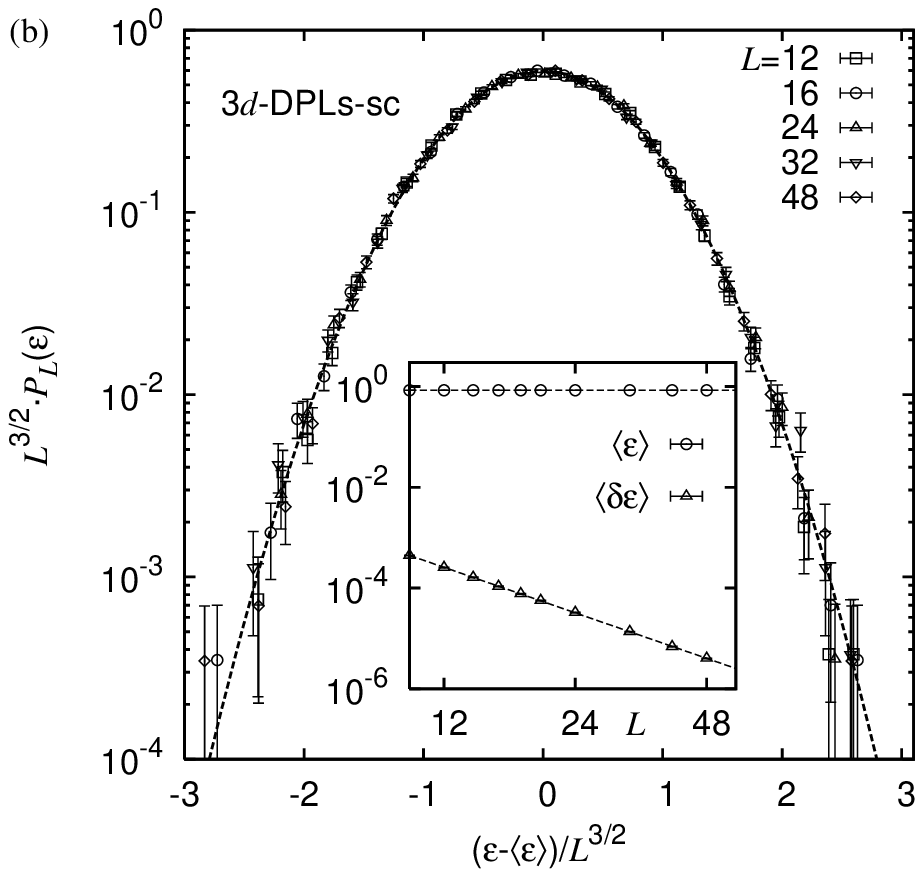}}
\caption{ \label{fig:fig7ab}
Results obtained for the pdf $P_L(\epsilon)$ 
of the configurational energy per lattice site $\epsilon$.
(a) Data obtained for DPLs on the $2d$ square lattice, 
(b) data for DPLs on the $3d$ simple cubic lattice.
In either case, the shape of $P_L(\epsilon)$ 
compares well to a Gaussian distribution (indicated by the 
dashed line). The insets illustrate the average and 
the instance to instance fluctuation of $\epsilon$ as function of
the system size $L$.
}
\end{figure}  

\paragraph{Number densities:}
First, we consider the scaling behavior of the number density $n(R_s)$ of small loops 
with size $R_s$. 
Upon analysis of the $2d$ and $3d$ data, we obtain scale invariant expressions of the 
form $n(R_s)\sim R_s^{-(d+1)}$ (not shown). 
Argumenting as above, it should not matter whether we consider the number density of 
loops with given spanning length or roughness instead of $n(R_s)$. 
In this regard, considering only non-spanning loops, we find $n(r)\sim r^{-3.00(3)}$ ($2d$-sq DPLs) and 
$n(r)\sim r^{-3.95(6)}$ ($3d$-sc DPLs), see inset of Figs.\ \ref{fig:fig4}(a),(b), 
where in either case the algebraic decay is governed by an exponent in good agreement 
with $(d+1)$. 
Note that, if the system of loops is scale invariant, the above scaling behavior can 
be inferred from a dimensional analysis \cite{vachaspati1984}.

Now, the scale invariance of the above number densities, together with the self similar 
scaling of the loop length, can be used to obtain the scaling assumption 
$n(\ell)\sim \ell^{-\tau}$ for the number density of loops with length $\ell$, 
see Refs. \cite{vachaspati1984,allega1990,hindmarsch1995}, defining the 
exponent $\tau\!=\!1+d/d_f$.
To illustrate this it is useful to consider the number of loops with size from 
$R_s$ to $R_s\!+\!dR_s$, given by $dN\!=\!n(R_s)~dR_s$. 
Using the leading order of the relation between the 
loop length and loop size, i.e. $\ell\!\sim\!R_s^{d_f}$ and hence 
$dR_s\!\sim\!\ell^{-1+1/d_f}~d\ell$, one can transform $dN$ by a change of 
variables $R_s\to\ell$ to yield
\begin{align*}
dN\sim R_s^{-(d+1)} ~dR_s \sim \ell^{-(1+d/d_f)}~d\ell \sim n(\ell)~d\ell.
\end{align*}

The scaling assumption for $n(\ell)$ above fits the data quite well, where, for the setups considered here, 
we yield the estimates of $\tau$ listed in Tab.\ \ref{tab:tab2}. The data for the 
$2d$-sq and $3d$-sc DPLs is shown in Figs.\ \ref{fig:fig4}(a),(b), respectively.
Finally, note that the values of $\tau$ and $d_f$ listed in Tab.\ \ref{tab:tab2} where 
obtained independently and agree with the scaling relation above within errorbars. 

Note that, for the $3d$-sc DPL setup, if one considers all loops (not only the non-spanning loops)
in order to compute the distribution $n(\ell)$, one finds a crossover to the scaling 
behavior $n(\ell)\!\sim\!\ell^{-1}$ in the limit of long loops (i.e.\ $\ell\!>\!10^3$ for 
the $3d$-sc lattice with $L\!=\!48$, see Fig.\ \ref{fig:fig4}(b)).
This behavior is a direct consequence of the periodic boundary conditions that allow
for long loops that wind around the lattice, i.e.\ those loops with a spanning index $j\!>\!0$. 
Such a crossover behavior was also observed within simulations of other string bearing models
\cite{allega1989a,allega1989b}, and, by analytic means, for a simple model of string formation on 
periodic lattices \cite{austin1994}.

\subsection{Energetic properties}\label{subsec:Results_erg}

\paragraph{Scaling of the configurational energy:}
The disorder averaged total weight $\langle\mathcal{E}\rangle$ of the 
loop configuration is an extensive quantity for which we found a clear 
asymptotic scaling that allows us to estimate
numerical values for $\langle \epsilon \rangle\!\equiv\!\langle|\mathcal{E}|/L^d\rangle$ 
as listed in Tab.\ \ref{tab:tab2}.
Here, we found no clear-cut deviations from the scaling behavior that 
might allow for an analysis of corrections to scaling.
In this regard, for the smallest $2d$-sq DPL system considered we found 
$\epsilon\!=\!0.5965$ ($L\!=\!12$), while for the largest system we obtained 
$\epsilon\!=\!0.5958$ ($L\!=\!256$). For the $3d$-sc DPL setup we similarly find
$\epsilon\!=\!0.84447$ ($L\!=\!10$) and $\epsilon\!=\!0.84446$ ($L\!=\!48$).

Moreover, the pdf $P_L(\epsilon)$ compares well to a Gaussian distribution,
as shown for $2d$-sq and $3d$-sc DPLs in Figs.\ \ref{fig:fig7ab}(a),(b), 
respectively. We further analyzed the instance to instance fluctuations 
$\langle\delta\epsilon\rangle=[ \langle\epsilon^2\rangle - \langle\epsilon\rangle^2 ]^{1/2}$ 
(inset of Figs.\ \ref{fig:fig7ab}(a),(b)),
for which we estimate $\langle\delta\epsilon\rangle\!=\!0.680(4)L^{-0.999(2)}$ ($2d$-sq DPLs)
and $\langle\delta\epsilon\rangle\!=\!0.68(1)L^{-1.504(6)}$ ($3d$-sc DPLs).  
We therefore deduce the asymptotic scaling $\langle\delta\epsilon\rangle \! \sim \!L^{-d/2}$.
Note that in terms of the extensive configurational energy $\mathcal{E}$, the latter reads
$\langle \delta \mathcal{E} \rangle \! \sim \! L^{d/2}$ and is thus in agreement with the
result for finite dimensional spin glasses \cite{wehr1990}.
As evident from Fig.\ \ref{fig:fig7ab}, the data sets for different system sizes show a nice 
data collapse if rescaled according to 
$P_L(\epsilon)=L^{-d/2} p_1[(\epsilon-\langle \epsilon \rangle)/L^{d/2}]$.

Considering DPLs for $2d$ square and honeycomb lattice graphs, one might
speculate whether the respective estimates of $\langle\epsilon\rangle$ reduce
to a unique value, apart from purely geometric factors. In this regard, 
assuming that $\epsilon$ depends on the fraction of occupied lattice sites
$\rho_{\rm cov}$ as well as the coordination number $z$ that characterizes
the particular lattice geometry ($2d$-sq: $z\!=\!4$; $2d$-hy: $z\!=\!3$), 
we find that the simple expression $\langle\epsilon\rangle\!=\!0.182\cdot z \cdot \rho_{\rm cov}$ 
fits the respective values of $\langle\epsilon\rangle$ (listed in Tab.\ \ref{tab:tab2}) quite well.

\begin{figure}[t!]
\centerline{
\includegraphics[width=1.0\linewidth]{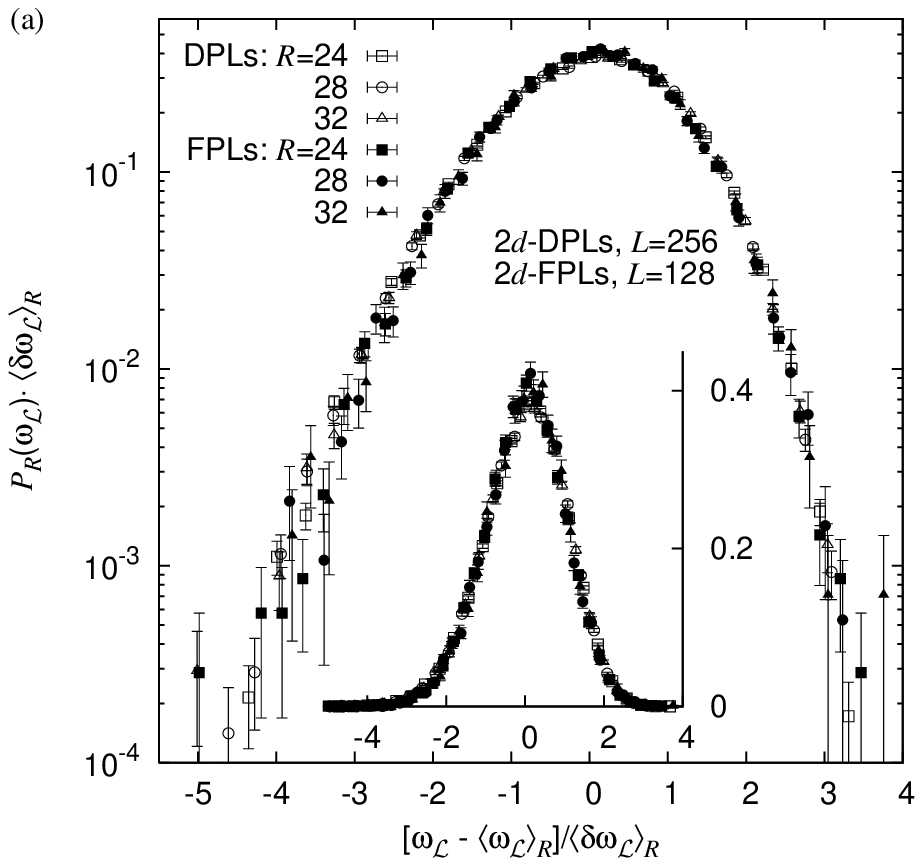}}
\centerline{
\includegraphics[width=1.0\linewidth]{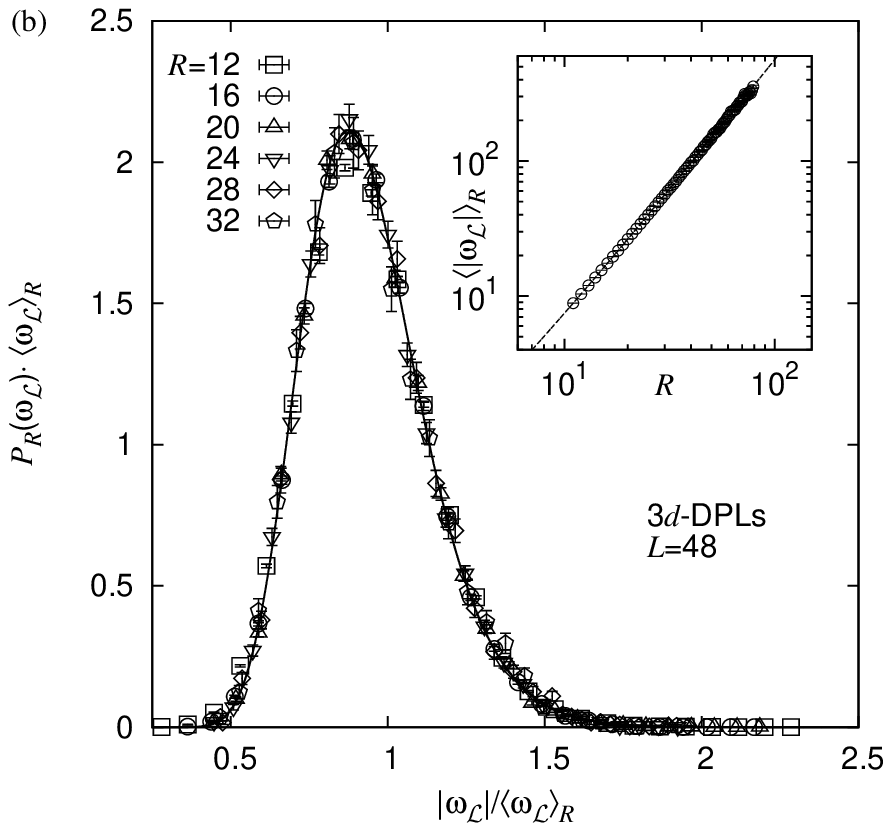}}
\caption{ \label{fig:fig5}
Scaling of the probability density function (pdf) $P_R(\omega_{\mathcal{L}})$
for the loop weight $\omega_{\mathcal{L}}$ of loops with spanning length $R$.
(a) the main plot shows the scaling of the pdf for $2d$ DPLs and FPLs on a 
semi-logarithmic scale, while the inset shows the same data on a linear scale.
(b) scaling behavior of the pdf for $3d$ DPLs, where the scaling function $\tilde{p}_2[|\omega_{\mathcal{L}}|/\langle\omega_{\mathcal{L}}\rangle_R]$ (solid line)
compares well to a log-normal distribution function as discussed in the text. The inset
shows the scaling of the average loop weight as function of the spanning length $R$.
}
\end{figure}  
\paragraph{Scaling properties of the loop weight:}
Further, we investigated the scaling of the average loop weight $\langle \omega_{\mathcal{L}} \rangle$ 
as function of the loop spanning length $R$ by considering the full ensemble of loops collected at
the largest system size simulated. In either case we find a good fit 
to the scaling form $\langle \omega_{\mathcal{L}} \rangle\!\sim\!R^\alpha (1+cR^{-\beta})$,
where we yield the exponents 
$\alpha\!=\!1.759(7)$ ($2.05(7)$) and $\beta\!=\!0.84(4)$ ($0.58(7)$) for $2d$-sq ($3d$-sc) DPLs
(results for the $3d$-sc data are shown in the inset of Fig.\ \ref{fig:fig5}(b)). 
For the $2d$-sq FPL setup and considering a similar scaling expression, we yield $\alpha\!=\!1.75(2)$ and
$\beta\!=\!0.87(17)$ (here, a more basic scaling function yields $\langle\omega_{\mathcal{L}}\rangle\!\sim\!(R+1.3(2))^{1.74(1)}$).
Note that a scaling according to the expression 
above, together with the numerical values of the respective exponents, indicates that 
$\langle\omega_{\mathcal{L}} \rangle\!\sim\!\ell$. 
Moreover, the root mean square deviation 
$\langle\delta\omega_{\mathcal{L}} \rangle\!=\![\langle\omega_{\mathcal{L}}^2\rangle-\langle\omega_{\mathcal{L}}\rangle^2]^{1/2}$
exhibits the same asymptotic scaling as the average loop weight, i.e.\ $\langle\delta\omega_{\mathcal{L}}\rangle\!\sim\!R^\alpha$
for sufficiently large values of $R$. However, for small values of $R$ there are also 
deviations from a pure power law scaling that might be accounted for by using a scaling 
expression of the form $\langle\delta\omega_{\mathcal{L}}\rangle\!\sim\!(R+\Delta R)^{\alpha^\prime}$.
E.g., for $2d$-sq DPLs (FPLs) we find $\Delta R\!=\!4(1)$ ($12.5(3)$) and 
$\alpha^\prime\!=\!1.75(2)$ ($1.77(2)$).

In either case, the pdf $P_R(\omega_{\mathcal{L}})$ of the loop weight for loops of different
spanning length $R$ exhibits the scaling behavior 
$P_R(\omega_{\mathcal{L}})\!\sim\!\langle\delta\omega_{\mathcal{L}}\rangle_R^{-1} p_2[(\omega_{\mathcal{L}}-\langle \omega_{\mathcal{L}} \rangle_R)/\langle\delta\omega_{\mathcal{L}}\rangle_R]$. 
Comparing $2d$-sq DPLs and FPLs, we find that 
both setups appear to have the same scaling function $p_2[\cdot]$, i.e.\ after rescaling 
both data sets displayed in Fig.\ \ref{fig:fig5}(a) fall onto a unique mastercurve. 

For the $3d$-sc case, we observe that 
$\langle\omega_{\mathcal{L}}\rangle_R\!\sim\!\langle\delta\omega_{\mathcal{L}}\rangle_R$
already for small values of $R$.
Thus, the data fits a slightly simpler scaling assumption, i.e.\
$P_R(\omega_{\mathcal{L}})\!=\!\langle\omega_{\mathcal{L}}\rangle_R^{-1} \tilde{p}_2[|\omega_{\mathcal{L}}|/\langle\omega_{\mathcal{L}}\rangle_R]$, 
where the scaling function $\tilde{p}_2[x]$ can further be parameterized by the 
log-normal distribution function
$\tilde{p}_2[x]\!=\!\exp[-(\log(x)\!-\!\mu)^2/(2 \sigma^2)]/(x \sigma \sqrt{2 \pi})$ with parameters 
$\mu\!=\!-0.089(1)$ and $\sigma\!=\!0.211(1)$, illustrated by the solid line in Fig.\ \ref{fig:fig5}(b).


\section{Conclusions \label{sect:conclusions}}
In the presented study, we performed simulations to characterize geometric and energetic properties
of densely and fully packed configurations of loops in the negative-weight percolation model. 
Therefore, we considered a Gaussian distribution of the edge weights and $2d$ square/honeycomb 
and $3d$ simple cubic lattice graphs. Further, so as to obtain configurations of minimum weight loops, 
we used a mapping of the NWP model to a combinatorial optimization problem that can 
be solved by means of exact algorithms.
In subsequent FSS analyses we observed that the scaling behavior of observables is best represented
by considering scaling expressions that also account for small power law corrections. 

Regarding the $2d$ setups, we found that the configurations of loops for both, DPLs and FPLs,
are characterized by the same values of the geometric scaling exponents $d_f$ and $\tau$. 
Moreover, the numerical value of $d_f$ found here coincides with the one measured for 
the smart kinetic walk (SKW) \cite{weinrib1985}. A loop-forming version of the SKW in two 
dimensions traces out the external perimeter of critical percolation clusters and 
was also used to model ring polymers at the $\theta^\prime$-point (which presumably is in the same
universality class as the $\theta$-point) on the honeycomb lattice \cite{coniglio1987}.
Also, it compares well to experimental results on the scaling of the radius of gyration of
$2d$ polymeric chains \cite{vilanove1980}. In that study, polymethylmetacrylate (PMMA) was 
spread onto an air-water interface at $16.5 ^\circ C$ that acted as a $\theta$-solvent for 
the respective polymer material. 
Further, the exponents $d_f$ and $\tau$ agree with those measured for fully packed loops 
on two discrete interface models termed ``random manifold'' and ``random elastic medium''
\cite{zeng1998}.
Finally, upon analysis of the pdfs that describe the scaling of the loop length as well 
as the loop weight, we found a universal behavior regarding the different $2d$ setups considered.

For the $3d$ setup, we observed two phases of loops containing ``small'' non-spanning loops 
and ``large'' spanning loops, respectively. Both phases are characterized by different geometric 
properties. 
Such a phenomenon was also observed for other string-bearing models, as e.g.\ cosmic string
networks \cite{allega1989a,allega1989b} or more general models of string formation on 
periodic lattices \cite{austin1994,allega1990}.
Regarding the small loop phase we conclude with a fractal dimension
$d_f\!\approx\!2.01(4)$. As discussed earlier, this scaling exponent indicates that the upper 
critical dimension for the DPLs takes the value $d_u^{\rm DPL}\!=\!3$. 
In contrast to this, earlier studies revealed that the upper critical dimension of the 
NWP model at the critical point where percolating loops first appear in the limit of large 
system sizes is $d_u\!=\!6$ \cite{melchert2010a}.
As for usual self-avoiding lattice curves for $d\!\geq\!d_u^{\rm DPL}$, one can further expect 
to find $d_f\!=\!2$. This we confirmed for DPLs on $4d$ hypercubic lattice graphs, where
we obtained the estimate $d_f\!=\!2.00(5)$.
Further, the large loop phase is a direct consequence of the periodic boundary conditions of the
simulated systems. Upon analysis, we found that approximately $90\%$ of the sites on the 
$3d$ simple cubic lattice are visited by loops and large loops comprise about $91\%$ of the 
loop segments on the lattice. Hence, these loops fill extensive parts of the ``simulation box'' 
and are strongly affected by the finite size of the accessible volume. 
Within the large loop phase we found evidence that the scaling properties of the loops further 
depend on their precise topology.

Finally, note that the algorithmic procedure presented in section \ref{sect:model} is
not limited to a particular disorder distribution or lattice setup and thus allows for the analysis 
of configurations of DPLs and FPLs for a multitude of lattice geometries and lattice dimensions.


\begin{acknowledgement}
OM acknowledges financial support from the VolkswagenStiftung (Germany)
within the program ``Nachwuchsgruppen an Universit\"aten''. 
The simulations were performed at the GOLEM I cluster for scientific 
computing at the University of Oldenburg (Germany).
\end{acknowledgement}


\bibliographystyle{unsrt}
\bibliography{literature_dpl.bib}

\begin{thebibliography}{10}

\bibitem{kardar1987}
M.~Kardar and Y.~C. Zhang.
\newblock {Scaling of Directed Polymers in Random Media}.
\newblock {\em Phys. Rev. Lett.}, 58:2087, 1987.

\bibitem{derrida1990}
B.~Derrida.
\newblock {Directed polymers in a random medium}.
\newblock {\em Physica A}, 163:71, 1990.

\bibitem{grassberger1993}
P.~Grassberger.
\newblock {Recursive sampling of random walks: self-avoiding walks in
  disordered media}.
\newblock {\em J. Phys. A}, 26:1023, 1993.

\bibitem{buldyrev2006}
S.~V. Buldyrev, S.~Havlin, and H.~E. Stanley.
\newblock {Optimal paths in strong and weak disorder: A unified approach}.
\newblock {\em Phys. Rev. E}, 73, 2006.

\bibitem{pfeiffer2002}
F.~O. Pfeiffer and H.~Rieger.
\newblock {Superconductor-to-normal phase transition in a vortex glass model:
  numerical evidence for a new percolation universality class}.
\newblock {\em J. Phys.: Condens. Matter}, 14:2361, 2002.

\bibitem{pfeiffer2003}
F.~O. Pfeiffer and H.~Rieger.
\newblock {Critical properties of loop percolation models with optimization
  constraints}.
\newblock {\em Phys. Rev. {\bf E}}, 67(5):056113, 2003.

\bibitem{vachaspati1984}
T.~Vachaspati and A.~Vilenkin.
\newblock {Formation and evolution of cosmic strings}.
\newblock {\em Phys. Rev. D}, 30(10):2036, 1984.

\bibitem{scherrer1986}
R.~J. Scherrer and J.~A. Frieman.
\newblock {Cosmic strings as random walks}.
\newblock {\em Phys. Rev. D}, 33, 1986.

\bibitem{hindmarsch1995}
H.~Hindmarsch and K.~Strobl.
\newblock {Statistical properties of strings}.
\newblock {\em Nucl. Phys. {\bf B}}, 437:471, 1995.

\bibitem{cieplak1994}
M.~Cieplak, A.~Maritan, and J.~R. Banavar.
\newblock {Optimal paths and domain walls in the strong disorder limit}.
\newblock {\em Phys. Rev. Lett.}, 72:2320, 1994.

\bibitem{melchert2007}
O.~Melchert and A.~K. Hartmann.
\newblock {Fractal dimension of domain walls in two-dimensional Ising spin
  glasses}.
\newblock {\em Phys. Rev. B}, 76:174411, 2007.

\bibitem{schwarz2009}
K.~Schwarz, A.~Karrenbauer, G.~Schehr, and H.~Rieger.
\newblock {Domain walls and chaos in the disordered SOS model}.
\newblock {\em J. Stat. Mech.}, 2009:P08022, 2009.

\bibitem{schwartz1998}
N.~Schwartz, A.~L. Nazaryev, and S.~Havlin.
\newblock {Optimal path in two and three dimensions}.
\newblock {\em Phys. Rev. E}, 58:7642, 1998.

\bibitem{rieger2003}
Heiko Rieger.
\newblock {Polynomial combinatorial optimization methods for analysing the
  ground states of disordered systems}.
\newblock {\em J. Phys. A}, 36(43), 2003.

\bibitem{SG2dReview2007}
A.~K. Hartmann.
\newblock Domain walls, droplets and barriers in two-dimensional ising spin
  glasses.
\newblock In Janke W, editor, {\em Rugged Free Energy Landscapes}, pages 67 --
  106, Berlin, 2007. Springer.

\bibitem{stauffer1979}
D.~Stauffer.
\newblock {Scaling theory of percolation clusters}.
\newblock {\em Phys. Rep.}, 54(1):1--45, 1979.

\bibitem{stauffer1994}
D.~Stauffer and A.~Aharony.
\newblock {\em {Introduction to Percolation Theory}}.
\newblock Taylor and Francis, London, 1994.

\bibitem{allega1990}
A.~M. Allega, L.~A. Fern\'andez, and A.~Taranc\'on.
\newblock {Configurational statistics of strings, fractals and polymer
  physics}.
\newblock {\em Nucl. Phys. B}, 332:760, 1990.

\bibitem{austin1994}
D.~Austin, E.~J. Copeland, and R.~J. Rivers.
\newblock {Statistical mechanics of strings on periodic lattices}.
\newblock {\em Phys. Rev. D}, 49:4089, 1994.

\bibitem{melchert2008}
O.~Melchert and A.~K. Hartmann.
\newblock {Negative-weight percolation}.
\newblock {\em New. J. Phys.}, 10:043039, 2008.

\bibitem{apolo2009}
L.~Apolo, O.~Melchert, and A.~K. Hartmann.
\newblock {Phase transitions in diluted negative-weight percolation models}.
\newblock {\em Phys. Rev. E}, 79:031103, 2009.

\bibitem{melchert2010a}
O.~Melchert, L.~Apolo, and A.~K. Hartmann.
\newblock {Upper critical dimension of the negative-weight percolation
  problem}.
\newblock {\em Phys. Rev. E}, 81(5):051108, 2010.

\bibitem{zeng1998}
C.~Zeng, J.~Kondev, D.~McNamara, and A.~A. Middleton.
\newblock {Statistical Topography of Glassy Interfaces}.
\newblock {\em Phys. Rev. Lett.}, 80(1), 1998.

\bibitem{ahuja1993}
R.~K. Ahuja, T.~L. Magnanti, and J.~B. Orlin.
\newblock {\em {Network Flows: Theory, Algorithms, and Applications}}.
\newblock Prentice Hall, 1993.

\bibitem{cook1999}
W.~Cook and A.~Rohe.
\newblock {Computing minimum-weight perfect matchings}.
\newblock {\em INFORMS J. Computing}, 11:138--148, 1999.

\bibitem{opt-phys2001}
A.~K. Hartmann and H.~Rieger.
\newblock {\em Optimization Algorithms in Physics}.
\newblock Wiley-VCH, Weinheim, 2001.

\bibitem{melchertThesis2009}
O.~Melchert.
\newblock {\em {PhD thesis}}.
\newblock not published, 2009.

\bibitem{practicalGuide2009}
A.~K. Hartmann.
\newblock {\em {Practical Guide to Computer Simulations}}.
\newblock World Scientific, Singapore, 2009.

\bibitem{comment_cookrohe}
For the calculation of minimum-weighted perfect matchings we use Cook and Rohes
  blossom4 extension to the Concorde library.

\bibitem{camarda2006}
M.~Camarda, F.~Siringo, R.~Pucci, A.~Sudb\o{}, and J.~Hove.
\newblock {Methods to determine the Hausdorff dimension of vortex loops in the
  three-dimensional $ XY $ model}.
\newblock {\em Phys. Rev. B}, 74(10):104507, 2006.

\bibitem{leibler1987}
S.~Leibler, R.~R.~P. Singh, and M.~E. Fisher.
\newblock {Thermodynamic behavior of two-dimensional vesicles}.
\newblock {\em Phys. Rev. Lett.}, 59:1989, 1987.

\bibitem{duplantier1990}
B.~Duplantier.
\newblock {Exact fractal area of two-dimensional vesicles}.
\newblock {\em Phys. Rev. Lett.}, 64(4):493, 1990.

\bibitem{allega1989a}
A.~M. Allega, L.~A. Fern\'andez, and A.~Taranc\'on.
\newblock {New regimes in the initial cosmic string network}.
\newblock {\em Phys. Lett. B}, 227:347, 1989.

\bibitem{allega1989b}
A.~M. Allega.
\newblock {Scaling in a string network}.
\newblock {\em Phys. Rev. D}, 40:1017, 1989.

\bibitem{wehr1990}
J.~{Wehr} and M.~{Aizenman}.
\newblock {Fluctuations of extensive functions of quenched random couplings}.
\newblock {\em J. Stat. Phys.}, 60:287, 1990.

\bibitem{weinrib1985}
A.~Weinrib and S.~A. Trugman.
\newblock {A new kinetic walk and percolation perimeters}.
\newblock {\em Phys. Rev. B}, 31:2993, 1985.

\bibitem{coniglio1987}
A.~Coniglio, N.~Jan, I.~Majid, and H.~E. Stanley.
\newblock {Conformation of a polymer chain at the theta' point: Connection to
  the external perimeter of a percolation cluster}.
\newblock {\em Phys. Rev. B}, 35:3617, 1987.

\bibitem{vilanove1980}
R.~Vilanove and F.~Rondelez.
\newblock {Scaling Description of Two-Dimensional Chain Conformations in
  Polymer Monolayers}.
\newblock {\em Phys. Rev. Lett.}, 45(18):1502, 1980.

\end{thebibliography}
\end{document}